\documentclass[12pt]{revtex4}

\usepackage{amsmath,amssymb,graphicx}
\usepackage{epsfig}
\usepackage{textcomp}
\usepackage{color}
\usepackage{cases}
\usepackage{epstopdf}
\usepackage{multirow}
\usepackage{hyperref}
\hypersetup{pdfpagemode=UseNone,colorlinks=true,linktoc=page,pdfborder={0 0 0},linkcolor=blue,citecolor=blue}
\usepackage{amstext}
\usepackage{bm}
\usepackage{color}

\begin{document}

\title{Large Fluctuations in Anti-Coordination Games on Scale-Free Graphs}

\author{Daniel Sabsovich$^{1,2}$, Mauro Mobilia$^{3}$, and Michael Assaf$^{1}$}
\affiliation{$^1$ Racah Institute of Physics, Hebrew University of Jerusalem, Jerusalem 91904, Israel}
\affiliation{$^2$ Department of Particle Physics \& Astrophysics, Faculty of Physics, Weizmann Institute of Science, POB 26, Rehovot, Israel}
\affiliation{$^3$ Department of Applied Mathematics, School of Mathematics,
  University of Leeds, Leeds LS2 9JT, U.K.}
\email{m.mobilia@leeds.ac.uk}
\email{michael.assaf@mail.huji.ac.il}

\begin{abstract}
We study the influence of the complex topology of  scale-free graphs
on the dynamics of anti-coordination games ({\it e.g.} snowdrift games). These
reference models are characterized by the coexistence (evolutionary stable mixed strategy) of two competing species, say
``cooperators'' and ``defectors'', and, in finite systems, by metastability and  large-fluctuation-driven fixation.
In this work, we use extensive computer simulations and an
effective diffusion approximation (in the weak selection limit) to determine under which circumstances, depending on
the individual-based update rules, the topology  drastically affects the long-time behavior of
anti-coordination games. In particular, we compute the variance of the number of cooperators in the metastable
state and the mean fixation time when the dynamics is implemented according to  the voter model (death-first/birth-second process)
and the link dynamics (birth/death or death/birth at random). For the voter update rule, we show that the scale-free
 topology effectively renormalizes the population size and as a result the statistics of observables depend on the network's degree
 distribution. In contrast, such a renormalization does not occur with the link dynamics update rule and we recover the same behavior as on  complete graphs.
\end{abstract}

\maketitle

\section{Introduction}
\label{sec:intro}

Evolutionary game theory (EGT) is a suitable framework to model the dynamics of populations
in which the success of one type depends on the actions of the others. In EGT, selection varies with the species
densities and is thus ``frequency dependent''~\cite{EGT1,EGT2,EGT3,Nowak}. This means that in the realm of EGT
the population composition and each species' fitness change continuously in time.
The ensuing evolutionary dynamics is commonly modeled deterministically in terms of the celebrated ``replicator equations'',
which are nonlinear differential equations~\cite{EGT1,EGT2,EGT3,Nowak,freqdepsel1,freqdepsel2} well-suited to describe very large populations.
The size of a real population is however always finite and is more realistically described by stochastic models
whose properties, such as demographic fluctuations, are known to often greatly influence the evolutionary
dynamics~\cite{weaksel1,weaksel2,weaksel3}. In particular, due to randomness, individuals of one species can take over and fixate the entire population. This stochastic phenomenon,
referred to as ``fixation'', is characterized by the fixation probability -- the probability that  a ``mutant type'' takes
over~\cite{Nowak,Kimura,Ewens} -- and the {\it mean fixation time} (MFT) which is the mean time for such an event to occur.
Population dynamics is also known to depend on the individuals' spatial arrangement: while a large body of work has focused on EGT
on spatially-homogeneous (well-mixed) populations~\cite{EGT1,EGT2,EGT3}, it is known that the outcome of the dynamics may be very different in
spatial settings, see {\it e.g.}~\cite{RPS1,RPS2,RPS3,RPS4,RPS5,RPS6}. For instance, spatial degrees of freedom have been found to promote cooperation in the prisoner's dilemma
game but to hinder coexistence in snowdrift games~\cite{Spatialcoop1,Spatialcoop2,Spatialcoop3,Xia15a,EGT1,EGT2,EGT3}.

 EGT was first introduced to study ecological dynamics~\cite{EGT1} and is also particularly well suited to model
the evolution of social behavior of interacting agents~\cite{EGT2,EGT3}. In this context, evolutionary games on
graphs~\cite{NetRef,Nets1,Nets2,Nets3} provides  a general unifying  framework that is able to capture the dynamics
of spatially structured  populations: it models how individuals interact with their neighbors, to reproduce or die, as prescribed by the underlying game
 ~\cite{EGTnets1,EGTnets2,EGTnets3,EGTgraphs,EGTgraphs-bis}. Of particular interest is the question of understanding how the
 network's topology affects the fixation properties of a given process, see {\it e.g.} Refs.~\cite{EGTgraphs,EGTgraphs-bis,Allen2016}. While it is difficult to give a general answer to
 this question~\cite{Ibsen15},  significant progress can be made when the selection pressure is weak, see
 {\it e.g.}~\cite{weaksel1,weaksel2,weaksel3}.
 In fact, most studies have focused on the biologically relevant and tractable limit of weak selection in which
 exact results have been obtained on regular graphs, see {\it e.g.}~\cite{EGTgraphs-bis,Taylor07,Chen13}.
 However,  much less is known about the fixation properties of evolutionary games on degree-heterogeneous graphs such as
 scale-free networks~\cite{NetRef,Nets1,Nets2,Nets3}. Most investigations on this important class of graphs
 have been carried out by means of computer simulations~\cite{Pacheco1,Pacheco2,hubs,Macie14,Xia15b},
 various approximation schemes~\cite{Tarnita09,Konno11,Allen2016}, and by considering special graphs~\cite{Broom11,Macie14}.
 The prisoner's dilemma game  has been extensively studied on scale-free networks and it has been found that
 the existence of nodes with high connectivity (hubs) can promote cooperation even under adverse
 conditions~\cite{Pacheco1,Pacheco2,EGTgraphs,hubs,Xia15b}, whereas the presence at the hubs of
 so-called facilitators~\cite{facil1,facil1bis}, which are agents that cooperate with cooperators and defect with defectors,
 tame the cooperation dilemma~\cite{facil2}. Recently, tools of statistical mechanics have been used to study simple
 evolutionary processes in which two types,  say {\it cooperators} and {\it defectors}, interact on degree-heterogeneous
 graphs under constant fitness and fixed (weak) selection
 pressure~\cite{VotNets31,VotNets32,VotNets11,VotNets12,VotNets13,VotNets21,VotNets22,VotNets23}.
 While these models  shed light on intriguing properties of evolution on complex graphs, such as the fact that it depends on
 the microscopic details of the update rules, they cannot describe evolutionary processes characterized by
 a {\it metastable} species coexistence prior to fixation like in the paradigmatic {\it anti-coordination games}
(ACGs)~\cite{EGT1,EGT2,EGT3,Gore09}, see {\it e.g.}~\cite{AM1,AM2,AM3,AM4,MA10,AM10,AM16}.
 In spite of the importance of systems like ACGs, their properties have been investigated mostly on regular
 lattices~\cite{Killingback96,Doebeli05,Spatialcoop2} and on small-world networks~\cite{Shang06,Qiu10}. In fact,
  there are very few results, mostly based on computer simulations~\cite{Pacheco1}, on how the scale-free
  topology affects the metastability and fixation properties of ACGs.
Here, the analysis attempted in Refs.~\cite{PRL12,AM13} is critically revisited and generalized.

In this work we study the joint effect of degree-heterogeneous topology and frequency-dependent
selection~\cite{freqdepsel1,freqdepsel2} on the dynamics of ACGs on scale-free networks under weak selection.
These are well-known EGT models, of particular importance in biology and ecology~\cite{EGT1,EGT2,EGT3,Gore09},
characterized by a long-lived coexistence state (metastability), and in which fixation is driven by large fluctuations~\cite{MA10,AM10}.
In particular, we investigate the influence of the microscopic update rule on the evolutionary dynamics. For the sake of concreteness,
here the individual-based dynamics is implemented according to two common update rules,
namely with the voter model (VM, death-first/birth-second process) and the link dynamics (LD, birth/death or death/birth at
random)~\cite{VotNets31,VotNets32,VotNets11,VotNets12,VotNets13,VotNets21,VotNets22,VotNets23}. By combining analytical  and
simulation means, we consider systems of large but finite size and determine the effect of the complex topology on the evolutionary dynamics. In particular, with the VM we show that the typical fluctuations in the number of cooperators in the metastable state is anomalous (their variance grows
superlinearly in the population size), and that the MFT has a stretched exponential dependence on the population size. We also show
that with the LD these quantities coincide in the leading order with the results on complete graphs. It is worth noting that
our approach is not limited to scale-free graphs, and is valid for general degree-heterogeneous networks, see, {\it e.g.}
Refs.~\cite{VotNets11,VotNets12,VotNets13,VotNets21,VotNets22,VotNets23}.

This paper is organized as follows: The  class of ACGs on networks that we consider is introduced in the next section.
Section III is dedicated to a description of the implementation of our computer simulations, while the analytical approach in terms of
 a multivariate diffusion theory  is presented in Sec.~IV. The various timescales that characterize the dynamics are presented in Sec.~V, where timescale separation is used to
derive an effective single-variate diffusion theory. Such an approach,  corroborated by extensive simulations of the individual-based system,
allows us to characterize the typical fluctuations in the metastable state by determining the variance in the number of cooperators in Sec.~VI,
and to obtain the MFT in Sec.~VII.  Finally, we summarize our findings and present our conclusions. Technical details about our
computational methods are provided in Appendix A and B.

\section{The Models: ACGs on complex networks}
\label{sec:model}
As in Ref.~\cite{PRL12}, we consider a scale-free network consisting of $N$ nodes
on which population dynamics takes place between two types of agents (see Appendix~\ref{app:AppendixA}), here referred to as
cooperators ($\textsf{C}$'s) and defectors ($\textsf{D}$'s). Each node is associated with
a binary  random variable: $\eta_i=1$ if the node  $i$ is occupied by a $\textsf{C}$,
whereas $\eta_i=0$ if it is occupied by a $\textsf{D}$.
The topology of the network is defined by its adjacency matrix
${\bm A}=[A_{ij}]$~\cite{NetRef,Nets1,Nets2,Nets3}, whose elements are $1$ if the nodes $ij$
are connected and $0$ otherwise, and its state (or population composition)
 is  described by $\{\eta_i\}^N=\{\eta_1, \dots, \eta_N\}$. The  underlying  complex graph is characterized by a degree distribution
 $n_k=N_k/N$, where $N_k$ denotes the number of nodes of degree $k$, whose $m^{th}$ moment is defined by
\begin{eqnarray}
\label{mu}
\mu_m\equiv \sum_k k^m n_k=\sum_i k_i^m /N.
\end{eqnarray}
Here $k_i$ is the degree of the node $i$, $\mu_1$ is the graph's mean degree and $N\mu_1/2$ is
the average number of links. There are standard methods to generate random networks whose nodes
are distributed according to a prescribed degree distribution, see {\it e.g.} Ref.~\cite{Nets1,Nets2,Nets3}. Here
we use the method outlined in Appendix~\ref{app:AppendixA} to generate the scale-free graphs
that we will consider in this work.
To study how the population composition changes in  time, we introduce  the density $\rho$ of cooperators in the network,
and the subgraph density   $\rho_k$ of $\textsf{C}$'s on nodes of degree $k$. These quantities are defined
by
\begin{equation}
\label{rho}
\rho\equiv \sum_{i} \eta_i/N\,,\;\;\;\;\;\;\rho_k \equiv \sum_{i}' \eta_i/N_k,
\end{equation}
where $\sum_{i}'$ denotes a summation restricted to the nodes $i$ of {\it fixed degree $k$}. We note that
$\rho=\sum_k n_k \rho_k$.
It is also useful to introduce the degree-weighted density of cooperators~\cite{VotNets11,VotNets12,VotNets13}:
\begin{eqnarray}
\label{omega}
\omega\equiv \frac{1}{N\mu_1}\sum_i k_i \eta_i=\sum_k \frac{k}{\mu_1} n_k \rho_k.
\end{eqnarray}

Here, we are interested in the class of anti-coordination games (ACGs) which are symmetric two-player two-strategy
games. According to the tenets of EGT~\cite{EGT1,EGT2,EGT3}, connected cooperators and defectors compete (or ``play'') pairwise according to
the payoff matrix
\begin{eqnarray}
\label{payoffM}
\begin{tabular}{c|c c}
vs & $\textsf{C}$  & $\textsf{D}$ \\
 \hline
$\textsf{C}$  &  $a$ & $b$  \\
$\textsf{D}$  & $c$ & $d$ \\
\end{tabular}
\end{eqnarray}
This specifies that two interacting cooperators get a payoff $a$, whereas defectors playing against each other
both get a payoff $d$. Moreover, when a $\textsf{C}$ plays against a $\textsf{D}$ the former's payoff is $b$
and the latter gets a payoff $c$. It is well known that various scenarios, including  ``cooperation dilemma'',
emerge depending on the
various values of  the entries of (\ref{payoffM})~\cite{EGT1,EGT2,EGT3,EGTnets1,EGTnets2,EGTnets3}. In this work, we focus on the important class of
ACGs for which $c>a$ and $b>d$. The class of ACGs includes the snowdrift game, for which
$c>a>b>d$, that is particularly relevant to biological applications, see {\it e.g.} Ref.~\cite{Gore09}.
\subsection{Well-mixed setting}
In the usual setting of EGT where the population is well-mixed, the expected payoffs (per individual) to cooperators and defectors
are respectively $\Pi^C(\rho)=a\rho+b(1-\rho)$ and $\Pi^D(\rho)=c\rho+d(1-\rho)$, while the population average payoff
is $\bar{\Pi}(\rho)=\rho \Pi^C(\rho) + (1-\rho)\Pi^D(\rho)$.
In such a setting, one of the main features of  ACGs is a coexistence state  in which a fraction $\rho^*$ of
$\textsf{C}$'s coexist with a density $1-\rho^*$ of
$\textsf{D}$'s  over a long period of time.  In the game-theoretic language the strategy $(\rho^*, 1-\rho^*)$ corresponds to playing cooperation $\textsf{C}$ with a frequency
$\rho^*$ and defection $\textsf{D}$ with  frequency $1-\rho^*$. This {\it mixed strategy} is known to be
evolutionary stable (but it is not a strict Nash equilibrium)~\cite{EGT1,EGT2,EGT3}. In order to find $\rho^*$, we write down the mean-field replicator equation (RE), which reads~\cite{EGT1,EGT2,EGT3,Nowak}:
\begin{equation}
\label{RE}
\frac{d}{dt}\rho(t)=\rho(t)(1-\rho(t))[\Pi^{C}(\rho(t))-\Pi^{D}(\rho(t))]=(a+d-b-c)\rho(t)(1-\rho(t))(\rho(t)-\rho^*).
\end{equation}
This equation is characterized by a stable interior fixed point
$\rho^*=(b-d)/(b+c-a-d)$ and unstable absorbing states $\rho=0$ (all-$\textsf{D}$) and $\rho=1$ (all-$\textsf{C}$). That is, in the deterministic picture, starting from any $0<\rho<1$ the system settles into the stable point $\rho^*$ and stays there forever. This picture, however, is altered when the  population size is finite ($N< \infty$), due to demographic fluctuations which ultimately drive the system into one of its two absorbing states. As a result, the stable fixed point in the language of the RE, $\rho^*$, becomes \textit{metastable}, and the probability to be in its vicinity slowly decays while the probability to be absorbed in either absorbing states slowly grows. It is well known that the mean decay time of the metastable state, which approximately equals the MFT, grows exponentially with $N$ on complete graphs (well-mixed population, $A_{ij}=1,\forall ij$)~\cite{MA10,AM10}.
\subsection{Spatially-structured setting}
When the population is spatially-structured and occupies the vertices of a graph,
the interactions are among nearest-neighbor  agents. The corresponding expected payoffs are thus
defined locally: If a node $j$ is occupied by a $\textsf{D}$ individual,
its neighbor $i$ receives a payoff $\Pi_{i}^C=b$ if the node $i$ is occupied by a
$\textsf{C}$ and the payoff of the agent node $i$ is $\Pi_{i}^D=d$ if it is a
$\textsf{D}$ individual.
The local reproductive potential, or fitness, of the agent at node $i$ whose neighbor $j$ is
a $\textsf{D}$ individual is proportional to the difference of their expected payoff relative to the population average
payoff $\bar{\Pi}_{i}(t)$ as perceived by the agent at node $i$. For the latter, we make the choice to consider
$\bar{\Pi}=\rho(t) \Pi_{i}^C +(1-\rho(t))\Pi_{i}^D$~\cite{PRL12}.
This  mean-field-like form reflects in a simple manner the fact that agents compare their payoffs with
 those of all others, leading to  metastability via a natural and analytically amenable mechanism.
As customary in EGT, we also introduce in the definition of the fitness a selection strength $s>0$, accounting for the interplay between demographic fluctuations and selection, as well as a baseline contribution, accounting for the chance contribution to the reproduction, which we set to $1$~\cite{Nowak,weaksel1,weaksel2,weaksel3,Kimura,Ewens}.
In this setting,  the fitnesses of a $\textsf{C}$ ($\textsf{D}$) player at node $i$
against a $\textsf{D}$ ($\textsf{C}$) player at a neighboring node $j$ are~\footnote{Naturally,
other update rules,
such as the Moran-like, the local-update  or Fermi-like rules, may also be considered. However,
for $s\ll 1$, all these update rules approximately coincide, see, {\it e.g.}, Refs.~\cite{weaksel2,weaksel3,facil1}.}
\begin{equation}
\label{fitCD}
f_{i}^C=1+s[\Pi_{i}^C-\bar{\Pi}_{i}]=1+s(b-d)(1-\rho)\,,\;\;\;\;f_{i}^D=1+s[\Pi_{i}^D-\bar{\Pi}_{i}]=1+s(c-a)\rho.
\end{equation}
We now proceed to specify the update rules by which, at an individual-based level, the population evolves.
Various types of update rules are possible and, in the case of models without fitness-dependent selection,
it has been found that the dynamics may depend crucially on the details of the underlying rules~\cite{VotNets11,VotNets12,VotNets13,VotNets21,VotNets22,VotNets23,VotNets31,VotNets32}.
Here, we consider two important choices: the so called (i) ``voter model'' (VM) rule~\cite{VotNets11,VotNets12,VotNets13}; (ii) and the
``link dynamics''
(LD)~\cite{VotNets31,VotNets32,PRL12,AM13}:
\\
(i) In the VM dynamics, a focal agent $i$ is chosen at random with probability $1/N$, and then one of its neighbors
is picked with a probability $1/k_i$. In this death-first/birth-second process, the focal agent dies and is replaced by
the picked neighbor with a probability proportional to the fitness of the latter. Or, equivalently when $0<s\ll 1$, and as implemented
in Ref.~\cite{VotNets11,VotNets12,VotNets13} for the biased VM and here in our simulations
(see below), the focal agent dies and is replaced by an offspring of
the picked neighbor with a probability proportional to the inverse of the focal agent's fitness, see  Sec.~\ref{sec:Impl}.
In practice, this means that with the VM at each time increment the population composition changes only when a neighboring
\textsf{C}\textsf{D} or \textsf{D}\textsf{C} pair interacts. The reactions $\textsf{C}\textsf{D} \to \textsf{D}\textsf{D}$
(death and replacement of \textsf{C} at $i$ by a \textsf{D}) and $\textsf{D}\textsf{C} \to \textsf{C}\textsf{C}$
(death and replacement of a \textsf{D} at $i$ by a \textsf{C}) thus occur with rates  given
 by the inverse of the individual \textsf{C}'s and \textsf{D}'s fitness respectively, i.e. with rates
\begin{equation}
 \label{rates_VM}
1/f^{C}=1-s(b-d)(1-\rho)+{\cal O}(s^2)\,,\;\;\;\;\;1/f^D=1-s(c-a)\rho+{\cal O}(s^2).
\end{equation}
\\
(ii) In the LD, a link is randomly selected at each time step and if it connects a
\textsf{C}\textsf{D} pair, one of the neighbors is randomly selected for reproduction with a rate
proportional to its fitness, while the other is replaced by the newly produced offspring.
In practice, this means that with the LD at each time increment the population composition changes only when a
\textsf{C}\textsf{D} or \textsf{D}\textsf{C} pair at neighboring nodes interact. Thus, the
reactions $\textsf{D}\textsf{C} \to \textsf{C}\textsf{C}$ and $\textsf{C}\textsf{D} \to \textsf{D}\textsf{D}$
 occur with rates given by the fitness $f^{C/D}$ of the agent \textsf{C} and \textsf{D} respectively, i.e. with rates
\begin{equation}
 \label{rate_LD}
f^{C}=1+s(b-d)(1-\rho)\,,\;\;\;\;\;f^D=1+s(c-a)\rho.
\end{equation}

In what follows, we consider the evolution of the ACGs under both VM and LD update rules and
we show that markedly different behavior emerges.

\section{Computer simulations of the evolutionary dynamics}
\label{sec:Impl}

 Before theoretically analyzing the evolution of ACGs with the
VM and LD, in this section we describe how the evolutionary dynamics
with the VM and LD have been implemented in our computational individual-based simulations.

 We begin by outlining the simulation of the evolution with the LD which has been performed using the Gillespie algorithm~\cite{Gillespie}. Our starting point is a scale-free network, see Appendix~\ref{app:AppendixA}, where each node is in one of two states - \textsf{C} (cooperator) or \textsf{D} (defector).
The following steps are repeated until fixation of either \textsf{C} or \textsf{D} occur (or until a prescribed
maximal number of time-steps have been performed):
\begin{enumerate}
\item Compute the density $\rho$ of \textsf{C} and the fitnesses of \textsf{C} and \textsf{D} according to Eqs.~(\ref{fitCD}).
\item Draw random numbers $R_1$ and $R_2$ uniformly distributed between $0$ and $1$.
\item Pick a random node $i$ and a random neighbor $j$ of $i$. If the states of $i$ and $j$ are different, then if $R_1<\frac{f^{C}}{f^{C}+f^{D}}$, both nodes become \textsf{C}. Otherwise, they become \textsf{D}.
\item Increment time by $\Delta t=-\frac{1}{N}\frac{\ln R_2}{f^{C}+f^{D}}$.
\end{enumerate}

The implementation of the VM case goes along the same lines as that of the LD model, except for the update of the randomly chosen node $i$ and its neighbor $j$ in step $3$. For the VM, if node $i$ is a \textsf{C} and node $j$ is a \textsf{D}, then $i$ becomes a \textsf{D} with probability $1/f^C$. Otherwise, if node $i$ is a \textsf{D} and node $j$ is a \textsf{C}, then $i$ becomes a \textsf{C} with probability $1/f^D$.

\section{Diffusion theory}
We are particularly interested in the biologically-relevant regime of weak selection strength.
In such a limit where $0<s\ll 1$~\cite{weaksel1,weaksel2,weaksel3,Kimura,Ewens}, the evolutionary dynamics is generally well
described in terms
of the so-called diffusion theory using the of appropriate forward and backward
Fokker-Planck equations (FPEs)~\cite{Gardiner}. Below, the multivariate FPEs for the subgraph densities  are derived.
We introduce the following quantities
for the  evolution with the VM
\begin{eqnarray}
 \label{psiVM}
 \Psi_{ij}^{{\rm VM}}=(1- \eta_i) \eta_j /f^D \quad \text{and}\quad \Psi_{ji}^{{\rm VM}}=(1-\eta_j)\eta_i /f^C,
\end{eqnarray}
while for the LD we use the quantities
\begin{eqnarray}
 \label{psiLD}
 \Psi_{ij}^{{\rm LD}}=(1- \eta_i) \eta_j f^C \quad \text{and}\quad \Psi_{ji}^{{\rm LD}}=(1-\eta_j)\eta_i f^D,
\end{eqnarray}
where $(1- \eta_i)\eta_j$ and $(1- \eta_j)\eta_i$  are non-zero only when the nodes $ij$ are occupied by a pair
$\textsf{D}\textsf{C}$ and $\textsf{C}\textsf{D}$, respectively. In an infinitesimal time increment $\delta t=1/N$
the subgraph density  $\rho_k$ changes by $\pm \delta \rho_k=\pm 1/N_{k}$ according to a
birth-death process~\cite{Gardiner} defined respectively by the
transition rates
\begin{eqnarray}
\label{transition}
T^{+}(\rho_k)=\sum_{i}'\sum_j \frac{A_{ij}}{N{\cal Q}} \Psi_{ij}^{{\rm VM}/{\rm LD}} \quad
\text{and} \quad T^{-}(\rho_k)=\sum_{i}'\sum_j \frac{A_{ij}}{N{\cal Q}}
\Psi_{ji}^{{\rm VM}/{\rm LD}},
\end{eqnarray}
where, from the definition of the VM and LD, we have~\cite{VotNets11,VotNets12,VotNets13}
\begin{eqnarray}
\label{Q}
{\cal Q}=
\left\{ \begin{array}{ll}
         k_i & \mbox{for the VM},\\
        \mu_1 & \mbox{for the LD}.\end{array} \right.
\end{eqnarray}
Given an agent at node $i$, the probability
to pick one of its neighbors $j$  for an update
is $A_{ij}/(N{\cal Q})$, and the transition  $\eta_i \to 1-\eta_i$
occurs with probability $ \sum_j \frac{A_{ij}}{N{\cal Q}} \left[\Psi_{ij}^{{\rm VM}/{\rm LD}} +\Psi_{ji}^{{\rm VM}/{\rm LD}}\right]$ \cite{VotNets11,VotNets12,VotNets13,PRL12}.

To make analytical progress, we assume that the degrees of the nodes  of the
underlying  networks are uncorrelated, i.e. we consider
degree-uncorrelated heterogeneous graphs. As explained in Appendix~\ref{app:AppendixA},
such an assumption, which is exact for Molloy-Reed networks~\cite{Molloy},
is here valid because degree-correlations are essentially negligible for the scale-free graphs that we consider.
In the realm of this often called ``heterogeneous mean-field'' or ``degree-based mean-field''
approximation, see {\it e.g.} Refs~\cite{epidemicsRMP,Vespignani12} and references therein,
we therefore write $A_{ij}=k_i k_j/(N{\cal Q})$.

We now substitute (\ref{rate_LD}), (\ref{psiVM}), (\ref{psiLD}) and  (\ref{Q}) into (\ref{transition}), use the above heterogeneous
mean-field  approximation and the identity $\sum_{i}'N^{-1} \eta_i =N n_k\rho_k$. In the limit of $s\ll 1$, the transition rates, $T^{+} (\rho_k) \equiv T^+_k$ and $T^{-} (\rho_k)
\equiv T^-_k$, thus become
\begin{eqnarray}
\label{T_VM}
T^+_k=n_k \omega (1-\rho_k)\left[1-s\rho (c-a) \right]\,,\;\;\;\;T^-_k=n_k (1-\omega)\rho_k \left[1-s(1-\rho)(b-d)\right],
\end{eqnarray}
for the VM, while for the LD the transition rates are
\begin{equation}
\label{T_LD}
T^+_k= n_k \frac{k}{\mu_1} \omega (1-\rho_k)\left[1+s(1-\rho)(b-d)\right]\,,\;\;\;\;T^-_k= n_k \frac{k}{\mu_1} (1-\omega)\rho_k \left[1+s\rho (c-a)\right].
\end{equation}
To remind the reader, $\omega$ is the degree-weighted density of cooperators, and is given by Eq.~(\ref{omega}).
In the limit of weak selection intensity ($0<s\ll 1$), the  birth-and-death process defined
 by transition rates (\ref{T_VM}) or (\ref{T_LD})~\cite{MA10,AM10} is well described in terms of
 the multivariate forward and backward FPEs whose  generators are respectively~\cite{Gardiner,Kimura,Ewens}
\begin{eqnarray}
\label{Gf}
{\cal G}_{\rm f}(\{\rho_k\})&=&\sum_k \left[-\frac{\partial}{\partial \rho_k}\frac{(T^+_k\!-\!T^-_k)}{n_k}  +
\frac{\partial^2}{\partial \rho_k^2} \frac{(T^+_k\!+\!T^-_k)}{2Nn_k^2} \right], \\
\label{Gb}
{\cal G}_{\rm b}(\{\rho_k\})&=&\sum_k \left[\frac{(T^+_k\!-\!T^-_k)}{n_k} \frac{\partial}{\partial \rho_k} +
\frac{(T^+_k\!+\!T^-_k)}{2Nn_k^2} \frac{\partial^2}{\partial \rho_k^2} \right].
\end{eqnarray}

\section{Timescale separation and effective diffusion approximation}
Solving the multivariate FPEs associated with the generators (\ref{Gf}) and (\ref{Gb}) is a formidable task.
Fortunately, the analysis greatly simplifies in the weak selection limit ($0<s\ll 1$) on which we focus
our attention. This simplification stems from a separation of timescales  which allows us to
reduce the dynamics to that of an effective single-variate process~\cite{VotNets11,VotNets12,VotNets13,VotNets21,VotNets22,VotNets23,PRL12}.
Interestingly, timescale separation has also been used to simplify the analysis of the spread of epidemics
on degree-heterogeneous networks, see {\it e.g.}~\cite{Parra16}.
Here, we first examine the case of the VM update, and then critically revisit the case of the  LD discussed in Refs.~\cite{PRL12,AM13}.
`
\begin{figure}
\includegraphics[width=0.9\linewidth]{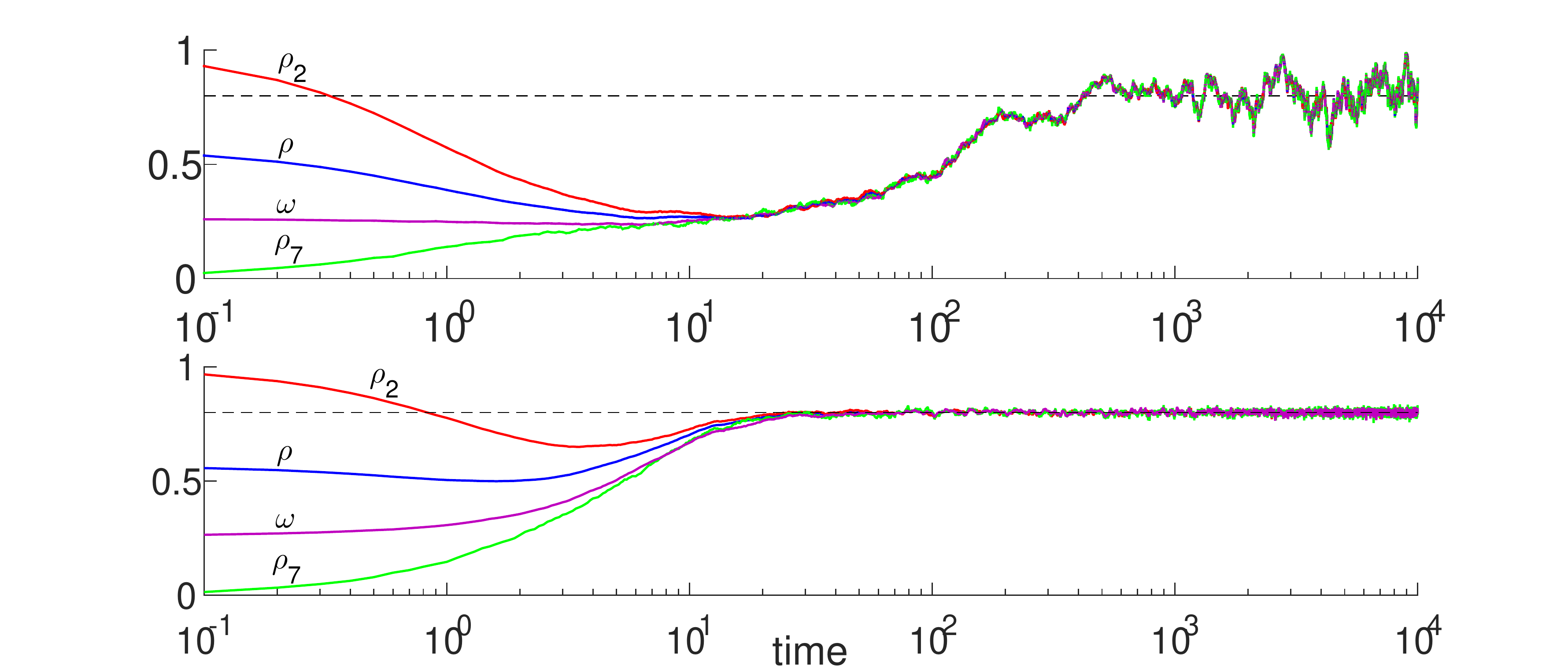}
\caption{(Color online).
Timescale separation with the VM update rule for weak (upper panel) and strong (lower panel) selection. Here, we illustrate
the dynamics of different density quantities in the system: the total density of the cooperators, $\rho$, the subgraph densities of the cooperators of degree $2$ and $7$, $\rho_2$ and $\rho_7$, and the degree-weighted density of cooperators, $\omega$, calculated by
summing up to degree $k = 200$.
In both panels we see a system with $N = 100, 000$ nodes and a power-law degree distribution with exponent
$\nu = 2.5$. The dynamics in both panels is defined by the  payoff matrix  (\ref{payoffM}) with entries $a = 1, b = 4, c = 1.75, d = 1$, with
weak selection, $s = 0.01$, in the upper panel and strong selection, $s = 1$, in the lower panel. The initial
state of the system is such that only nodes with degree $k \leq 3$ are cooperators, the others being defectors (at
$t = 0$: $\rho_2 = 1$ and $\rho_7= 0$). In both panels the total density $\rho$ and the subgraph densities $\rho_k$
converge to $\omega$ on a time scale of ${\cal O}(1)$ , and then $\omega$ converges to the stable interior fixed point
$\rho^* = 0.8$ (dashed line) on a timescale
of order  ${\cal O}(1/s)$. Fixation occurs, on a much larger timescale  $t\gg s^{-1}$ (not shown here), see text.
}
\label{Fig1}
\end{figure}
\subsection{Timescale separation \& effective diffusion theory for the VM update rule}

 For the dynamics with the VM update, the quantity  $\omega$ is conserved
 on a timescale
$t\ll s^{-1}$~\cite{VotNets11,VotNets12,VotNets13}. This result, illustrated in Fig.~\ref{Fig1}, can be understood by
computing the rate of change of $\delta  \bar{\omega}=\sum_i k_i (1-2\eta_i)/(N\mu_1)$\cite{VotNets11,VotNets12,VotNets13}:
$\delta\bar{\omega}/\delta t=\dot{\bar{\omega}} =\sum_{ij} A_{ij}[\bar{\psi}_{ij}-\bar{\psi}_{ji}]/(N\mu_1)$, where the bar
denotes the ensemble average. By using the mean-field approximation $A_{ij}=k_i k_j/(N{\cal Q})$, we
find that $\dot{\bar{\omega}}=(a + d-b-c)s\bar{\omega}(1-\bar{\omega})(\bar{\rho}-\rho^*)$. This means
that on a timescale $t\ll s^{-1}$ the quantity $\omega$ is approximately constant.

Furthermore, on such a timescale, the mean-field rate equation for the subgraph density $\bar{\rho}_k$ satisfies $\dot{\bar{\rho}}_k=(T_k^+ -T_k^-)/n_k=\bar{\omega}-
\bar{\rho}_k +  {\cal O}(s)$. Hence, after a timescale $t={\cal O}(1)$, all $\bar{\rho}_k$'s converge to $\omega$ and we have $\bar{\rho}_k \approx \bar{\omega}\approx$ constant. Thus, the subgraph densities become independent of $k$. Clearly this implies that $\rho=\sum_k \rho_k n_k\approx
\rho_k \approx \omega$. From the above, we therefore infer that when $t\gtrsim {\cal O}(1)$ we simply
have $\rho_k\approx
\rho \approx \omega$ and these quantities evolve together towards their common value $\rho^*$.

As confirmed in Fig.~\ref{Fig1}, at timescales $t\gtrsim {\cal O}(1)$
all the densities $\rho$ and $\rho_k$ converge to $\omega$, and reach the vicinity of $\rho^*$ after a timescale of ${\cal O}(s^{-1})$. This scenario lasts until a chance fluctuation eventually causes the fixation of either
$\textsf{C}$ or $\textsf{D}$ on a much longer time scale ($t\gg s^{-1}$).
In summary, as illustrated by Fig.~\ref{Fig1}, with the VM update
rule we distinguish three timescales: (i) at $t\gtrsim {\cal O}(1)$, $\omega$ is approximately constant and $\rho$ and $\rho_k$ converge to it; (ii) at $t\gtrsim {\cal O}(s^{-1})$, $\rho_k \approx \rho \approx \omega$ converge to $\rho^*$, and (iii) at $t\gg {\cal O}(s^{-1})$ fixation occurs.
The comparison of the top and bottom panels of  Fig.~\ref{Fig1}, illustrates that the separation of the three timescales
breaks down when $s$ becomes ${\cal O}(1)$. Overall, the effective diffusion approximation presented below
is valid only in the limit of weak selection, i.e. for $0<s\ll 1$.

Since we are interested in metastability and fixation, which occur when $\rho_k \approx \rho \approx
\omega$, we can legitimately approximate $\rho_k$ and $\rho$ by $\omega$~\cite{VotNets11,VotNets12,VotNets13}. We therefore substitute $\rho_k$
and $\rho$ by $\omega$ in the transition rates (\ref{T_VM}) and replace $\partial_{\rho_k}$ by $(k n_k/\mu_1)\partial_{\omega}$
in the generators (\ref{Gf}) and (\ref{Gb}). Under selection of weak intensity, $0<s\ll 1$,
this yields the single-variate effective forward and backward
generators:
\begin{eqnarray}
\label{Gf_eff}
\widetilde{{\cal G}}_{\rm f}(\omega)&=&
-\tilde{s}\frac{\partial}{\partial \omega} [\omega (1-\omega)(\rho^*-\omega)] +\frac{\mu_2}{N(\mu_1)^2}
\frac{\partial^2}{\partial \omega^2}[\omega (1-\omega)]
\\
\label{Gb_eff}
\widetilde{{\cal G}}_{\rm b}(\omega)&=&\omega (1-\omega)\left[\tilde{s}(\rho^*- \omega)\frac{\partial}{\partial \omega}
+\frac{\mu_2}{N(\mu_1)^2}
\frac{\partial^2}{\partial \omega^2}
\right],
\end{eqnarray}
where $\tilde{s}=(b+c-a-d)s$. Equations~(\ref{Gf_eff}) and (\ref{Gb_eff}) are among the main results of this paper. It has to be noted that in the diffusion term we have neglected the subleading contributions on the order of
${\cal O}(s)$ since we are working in the weak selection limit.

The effective diffusion theory therefore predicts that the main influence of the scale-free topology under the VM update rule
is to renormalize the population size $N$ into $N_{{\rm eff}} =N (\mu_1)^2/\mu_2$~\cite{VotNets11,VotNets12,VotNets13,VotNets21,VotNets22,VotNets23}.
Below we show that numerical simulations fully support this prediction, and we discuss how the
effective population size $N_{{\rm eff}} $ affects the metastability and fixation properties of the ACGs.
A relevant result for our analysis is the fact that for scale-free networks  the  degree distribution
$n_k \sim k^{-\nu}$ with $\nu>2$, such that $\mu_1$ is finite, holds up to the maximum degree estimated to scale as
$k_{{\rm max}}\sim N^{1/(\nu-1)}$\footnote{In principle, the algorithm that we have used  generates  scale-free graphs
and allows for the existence of nodes
of degree $k>k_{{\rm max}}$. Yet, as their number is negligible, these nodes play no statistical role, see Appendix A.
It is also worth  noting that we focus on scale-free graphs with exponent $\nu>2$ since these have a
finite mean degree $\mu_1$.} (size of the largest hub)~\cite{KS02}. Therefore,
the effective population size scales as~\cite{VotNets11,VotNets12,VotNets13}
\begin{eqnarray}
\label{Neff}
N_{{\rm eff}} =N \frac{(\mu_1)^2}{\mu_2}\sim
\left\{ \begin{array}{ll}
         N, &  \nu>3,\\
        N/\ln{N}, &  \nu=3,\\
        N^{\alpha}, & 2<\nu<3,
        \end{array} \right.
\end{eqnarray}
where we have introduced the exponent
\begin{eqnarray}
 \label{alpha}
 \alpha=\frac{2(\nu-2)}{\nu-1}.
\end{eqnarray}

\begin{figure}
\includegraphics[width=0.9\linewidth]{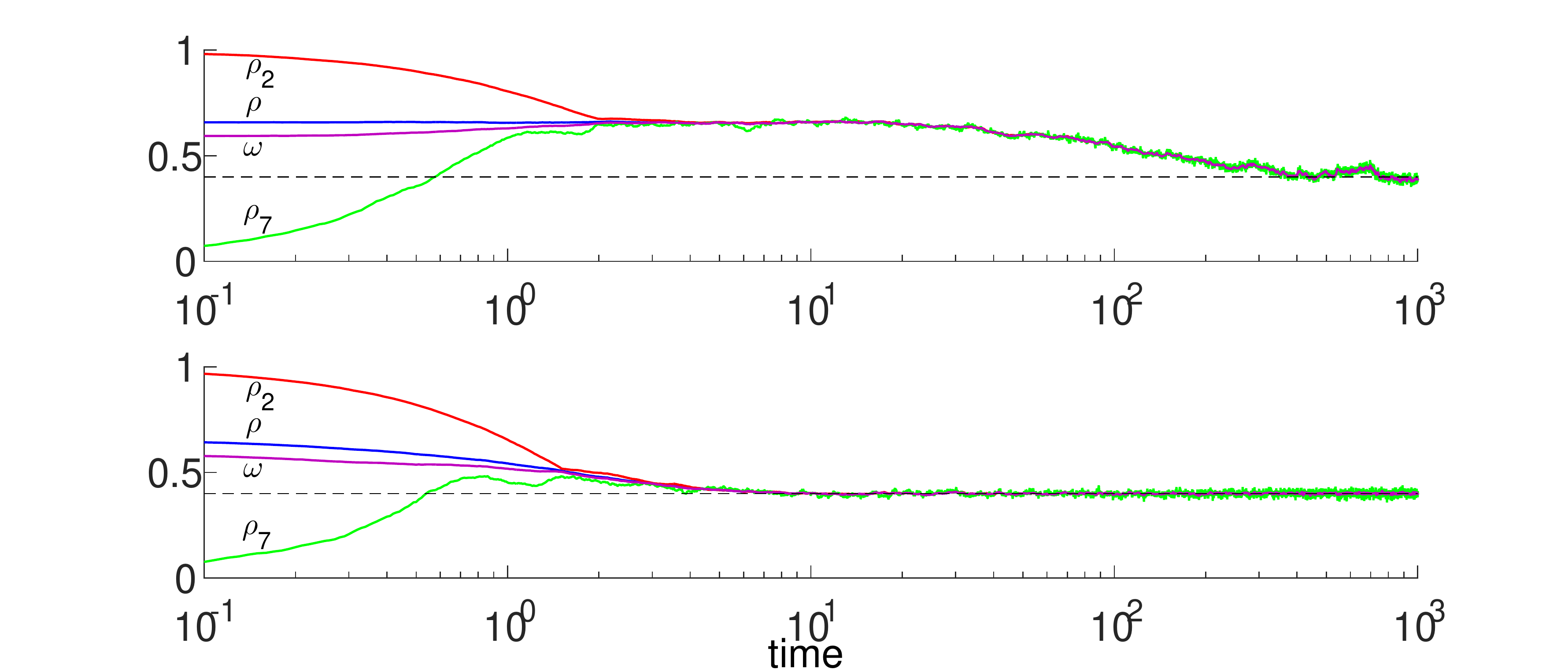}
\caption{(Color online).
Timescale separation with the LD update rule for weak (upper panel) and strong (lower panel) selection. Here, we illustrate
the dynamics of different density quantities in the system: the total density of the cooperators, $\rho$, the subgraph densities of the cooperators of degree $2$ and $7$, $\rho_2$ and $\rho_7$, and the degree-weighted density of cooperators, $\omega$, calculated by summing up to order $k=100$. In both panels we see
a system with $N=100,000$ nodes and power-law degree distribution with exponent $\nu=2.9$.
The dynamics in both panels is defined by the payoff matrix  (\ref{payoffM}) with entries $a=1,\,b=1.5,\,c=1.75,\,d=1$,
and selection strength $s=0.01$ (upper panel) and $s=1$ (lower panel). The
initial state of the system was such that  nodes of an even degree were cooperators, and nodes with an odd degree were defectors (at $t=0$: $\rho_{2}=1$ and $\rho_{7}=0$).
We can see in both cases that  $\omega$ and the subgraph densities $\rho_k$ converge to the total density $\rho$ on a time scale of $O\left(1\right)$,
and then $\rho$ converges to the stable interior fixed point $\rho^{*}=0.4$ (dashed line)  on a time scale of $O\left(1/s\right)$. Fixation occurs, on a much larger timescale  $t\gg s^{-1}$ (not shown here), see text.}
\label{Fig2}
\end{figure}
\subsection{Timescale separation \& effective diffusion theory for the LD update rule}

A similar reasoning holds also for the LD update rule. Yet, here it is crucial to realize
that the reference observable is the density of cooperators $\rho$. Indeed, under the LD
$\rho$ is conserved  on the time scale
$t\ll s^{-1}$~\cite{VotNets11,VotNets12,VotNets13,VotNets31,VotNets32} as one can check by noting that
$\delta \bar{\rho}= \sum_i(1-2\eta_i)/(N\mu_1)$ which, proceeding as above, yields
$\dot{\bar{\rho}} =\sum_{ij} A_{ij}[\bar{\psi}_{ij}-\bar{\psi}_{ji}]/(N\mu_1)
=(a + d-b-c)s\bar{\rho}(1-\bar{\rho})(\bar{\rho}-\rho^*)$.
 This indicates that $\rho$ remains constant when $t\ll s^{-1}$ and then
 relaxes to its metastable value $\rho^*$ on a timescale $t ={\cal O}(s^{-1})\gg 1$; this is corroborated by Fig.~\ref{Fig2}.

Furthermore, at the mean-field level one obtains
$\dot{ \bar{\rho}}_k=(T^+_k(\bar{\rho}_k)-T^-_k(\bar{\rho}_k))/n_k \sim
(\bar{\omega}- \bar{\rho}_k)k/\mu_1$, valid at times $t\gtrsim {\cal O}(1)$.
Thus, as illustrated in Fig.~\ref{Fig2}, this means that on a time scale of $t\gtrsim{\cal O}(1)$,
$\rho_k$ and $\omega$ approach $\rho$. Then, all these quantities evolve together towards their common
metastable value $\rho^*$, which is reached at a timescale of $t ={\cal O}(s^{-1})\gg 1$, and fluctuate around it before fixation
occurs at a much later stage, see below.

To study metastability and fixation in ACGs, we can therefore restrict our attention in the
regimes where  $\rho_k \approx \omega \approx  \rho$, and approximate $\rho_k$ and
$\omega$ by $\rho$~\cite{VotNets11,VotNets12,VotNets13}.
A one-body description of the dynamics is thus obtained by substituting $\rho_k$ and $\omega$ by $\rho$ in the
transition rates (\ref{T_LD}) and
by replacing $\partial_{\rho_k}$ by
$n_k\partial_{\rho}$
in the generators (\ref{Gf}) and (\ref{Gb}). This yields the single-variate effective forward and backward
generators, respectively:
\begin{equation}
\label{Gf_eff_LD}
\hspace{-5mm}
\widetilde{{\cal G}}_{\rm f}(\rho)=
\frac{\partial}{\partial \rho} \left[\tilde{s}\rho (1\!-\!\rho)(\rho\!-\!\rho^*) +\frac{1}{N}
\frac{\partial}{\partial \rho}\rho (1\!-\!\rho)\right],
\;\;
\widetilde{{\cal G}}_{\rm b}(\rho)=\rho (1\!-\!\rho)\left[\tilde{s}(\rho^*\!-\! \rho)\frac{\partial}{\partial \rho}
+\frac{1}{N}
\frac{\partial^2}{\partial \rho^2}
\right].
\end{equation}
It has to be noted that these generators are the same as those obtained in a
well-mixed population (on a complete graph) of size $N$; here, as opposed to Eqs.~(\ref{Gf_eff}) and (\ref{Gb_eff}), the  population size is not renormalized by the graph's topology. This is similar
 to what was found for models without selection
and with frequency-\textit{independent} selection evolving with the LD~\cite{VotNets11,VotNets12,VotNets13,VotNets21,VotNets22,VotNets23,VotNets31,VotNets32}.
As discussed in more detail below,  here we show that the metastability and fixation properties of ACGs evolving with the LD are the same in the leading order as those on a complete graph of size $N$. Therefore, the fixation probability and MFT of the ACGs with the LD
are not affected by the scale-free topology, contrary to what was reported in Ref.~\cite{PRL12,AM13}. Yet, a completely different scenario emerges with the VM update rule, where the scale-free topology strongly affects the long-time behavior.

\section{Fluctuations in the metastable state}
We can use the effective diffusion approximation to study the typical fluctuations in the long-lived
metastable state at times $t\gtrsim s^{-1}\gg 1$. In this regime,
the one-body  forward FPE   generators
(\ref{Gf_eff}) and (\ref{Gf_eff_LD}) allow us to describe the dynamics in the metastable state in
terms
of an Ornstein-Uhlenbeck process (OUP) obtained by linearizing the drift term about the metastable state
$\rho^*$ in  (\ref{Gf_eff}) and (\ref{Gf_eff_LD}) and by evaluating the diffusion term at $\rho^*$.

\begin{figure}
\includegraphics[width=0.90\linewidth]{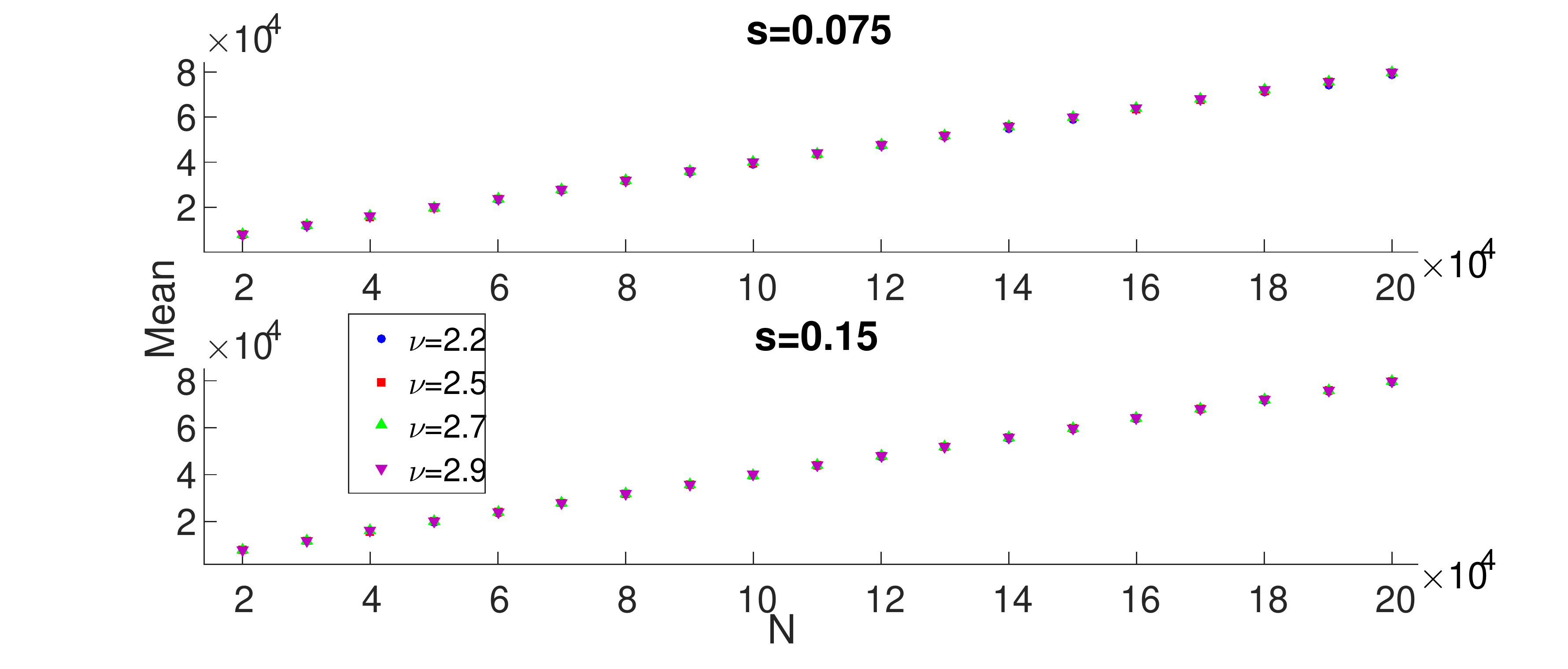}
\caption{(Color online).
Mean number of cooperators, $\left\langle N\rho\right\rangle $, with the VM update rule as
a function of the system size $N$: Results are shown for selection $s = 0.075$ (upper panel) and  $s = 0.15$ (lower panel), and for different
values of the exponent $\nu$ of the degree distribution $n_k$, see legend and Appendix \ref{app:AppendixB}. In the initial state of the system each node was a cooperator
with a $50\%$ probability. The dynamics is defined by the payoff
matrix (\ref{payoffM}) with entries $a=1,\,b=1.5,\,c=1.75,\,d=1$. Here, $\rho^*=0.4$ and the averaging was done from $t=500$  until $t=25000$
 (where we have omitted those simulations where fixation occurred within such a time window).
The errors are inside the range of the plotted data points.}
 \label{FigMean}
\end{figure}

\begin{figure}
\includegraphics[width=1.0\linewidth]{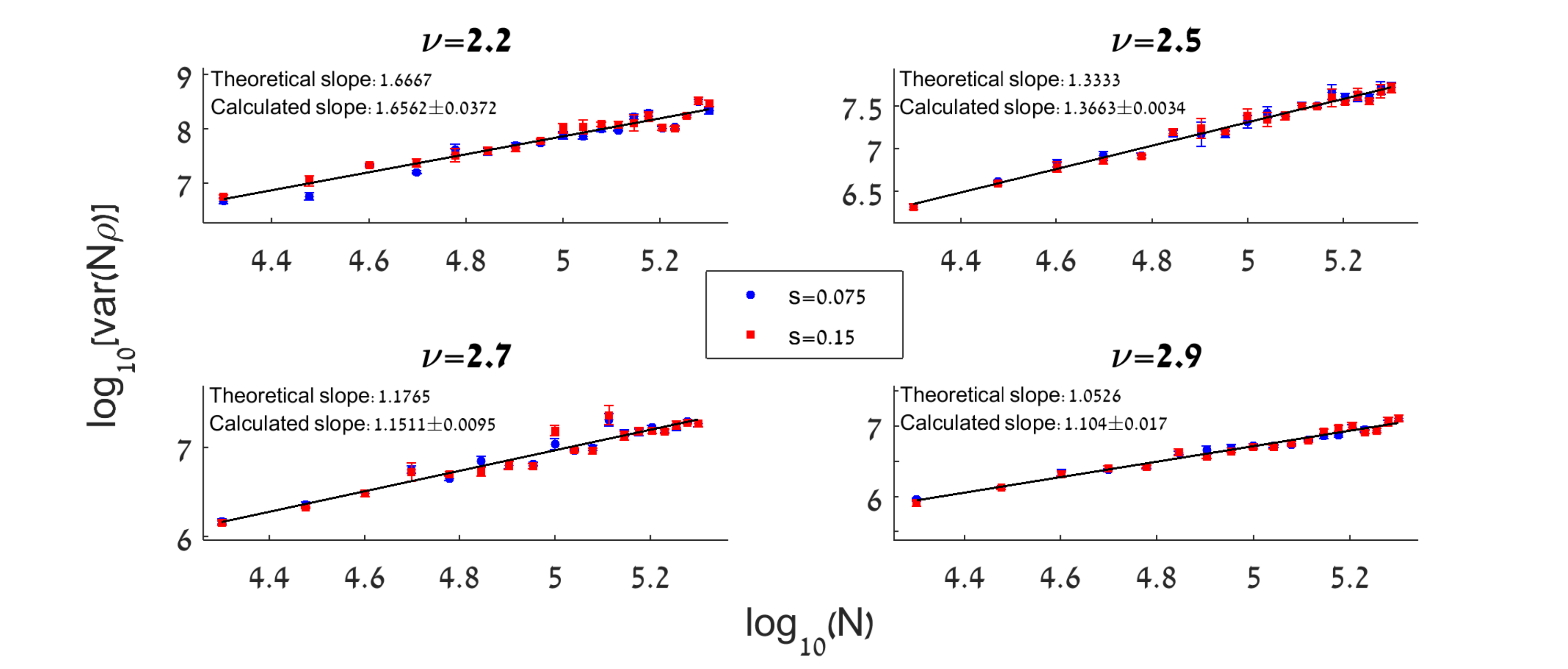}
\caption{(Color online). Variance of the number of cooperators, ${\rm var}\left(N\rho\right)$,  with the VM update rule as
a function of the system size $N$: Results are shown for $s=0.075$ and $s=0.15$, see legend, and different
values of $\nu$ of the degree distribution $n_k$, see panel titles and Appendix \ref{app:AppendixB}. The initial conditions and payoff matrix parameters (\ref{payoffM}) are the same as in Fig.~\ref{FigMean}.
The results are compared with the theoretical prediction $\log\left({\rm var}\left(N\rho\right)\right)\sim 2/\left(\nu-1\right)\log N$.
As seen in the plots, there is a good agreement between the measured data
and the theoretical prediction. For each subplot with a different
value of $\nu$, the legend reports the theoretical prediction
 $2/(\nu-1)$ for the slope, and the measured slope (averaged over those
obtained for  different values of $s$). Here to get the collapse between the lines with different $s$, the data with $s=0.15$ is
shifted by $\log 2$ since ${\rm var}\left(N\rho\right)\sim s^{-1}$. In addition, averaging was done
from $t=500$  until $t=25000$ (where we have omitted those simulations where fixation occurred within such a time window).}
 \label{FigVar}
\end{figure}

\subsection{Fluctuations in the metastable state with the VM update}
For ACGs evolving with the VM, in the realm of the effective  diffusion approximation,
the (forward) generator of the OUP for $\xi=\omega-\rho^*\approx \rho-\rho^*$
is therefore
\begin{eqnarray}
\label{Gf_VM_OU}
\widetilde{{\cal G}}_{\rm f}(\xi)=
\rho^* (1-\rho^*)\left[-\tilde{s} \frac{\partial}{\partial \xi} \xi
+ \frac{1}{N_{\rm eff}}  \frac{\partial^2}{\partial \xi^2}\right].
\end{eqnarray}
Here, we readily recognize the generator of an OUP with drift and  diffusion parameters
$\tilde{s}\rho^* (1-\rho^*)$ and $2\rho^* (1-\rho^*)/N_{\rm eff}$, respectively.
Using the well-known properties of the OUP~\cite{Gardiner}, we readily
verify that on average the deviations from the metastable state vanish:
$\langle \xi(t)\rangle= \xi(0)e^{-\tilde{s}\rho^* (1-\rho^*)t} \to 0$, where the
square bracket denotes the ensemble average over the probability density $p(\xi,t)$
of the forward FPE $(\partial_t -\widetilde{{\cal G}}_{\rm f}(\xi))p(\xi,t)=0$
[with zero-flux boundary conditions]. This means that, on average, the deviations from $\rho^*$ vanish very quickly in the
metastable state. Since  $\langle \xi(t)\rangle=0$, the mean number of cooperators in  the metasable state is
$\langle N\rho \rangle=N\rho^*$ and therefore grows linearly with the number of nodes $N$. This result is confirmed in Fig.~\ref{FigMean} (obtained from the analysis of  stochastic simulations data, see Appendix \ref{app:AppendixB}), showing that the mean number of cooperators in the metastable state increases linearly in $N$ with a slope $\rho^*$.

It is quite interesting to consider the variance of the variable
$\xi$. Since initially, ${\rm var}(\xi(0))=\langle \xi^2(0)\rangle=0$, we find at times $t\gtrsim s^{-1}\gg 1$
\begin{eqnarray}
\label{var_VM}
\langle \xi^2(t)\rangle=
\frac{1}{\tilde{s}N_{\rm eff}}\left[1-e^{-2\tilde{s}\rho^* (1-\rho^*)t}\right]\to \langle \xi^2\rangle=\frac{1}{\tilde{s}N_{\rm eff}}
\end{eqnarray}
where $N_{\rm eff}$ is given by Eq.~(\ref{Neff}).
Using~(\ref{var_VM}), one can find the variance of the number of cooperators in the metastable state:
${\rm var}(N\rho)=N^2(\langle \rho^2\rangle -\langle \rho\rangle^2)\approx
N^2(\langle \omega^2\rangle -(\rho^*)^2)\approx N^2\langle \xi^2\rangle\sim
N^2/(\tilde{s}N_{\rm eff})$. Hence, with the definition of $\alpha$ given by (\ref{alpha}), we find
\begin{eqnarray}
\label{var2}
{\rm var}(N\rho)=
N^2(\langle \rho^2\rangle - \langle \rho \rangle^2)  \sim
\left\{ \begin{array}{ll}
         N, &  \nu>3,\\
        N\ln{N}, &  \nu=3,\\
        N^{2-\alpha}=N^{2/(\nu-1)}, & 2<\nu<3.
        \end{array} \right.
\end{eqnarray}
In particular, we notice that for scale-free graphs with $2<\nu<3$ that are characterized by nodes of high
degree~\cite{NetRef}, the variance increases faster than linearly in the system size
since $2/(\nu-1)>1$. In other words, the effective diffusion theory predicts that
$\log{({\rm var}(N\rho))}\sim 2/(\nu-1)\log N$ when $2<\nu<3$.
This result is corroborated by extensive computer simulations outlined
in Appendix \ref{app:AppendixB} and illustrated in Fig.~\ref{FigVar}. Here, the theoretical prediction (\ref{var2})
is in very good agreement (within error bars) with the results of stochastic simulations over a broad range of values  $2<\nu<3$.
These results mean that the typical fluctuations in the metastable state on
scale-free graphs with exponent $2<\nu<3$ are much stronger than on complete graphs where the cooperator variance
 scales as $N$, see below. In Section VII, we explore how the strong fluctuations arising on scale-free networks with exponent  $2<\nu<3$ also dramatically affect the MFT.

\subsection{Fluctuations in the metastable state with the LD}
For ACGs evolving with the LD, in the realm of the effective  diffusion approximation,
the (forward) generator of the OUP for $\xi= \rho-\rho^*$
is again given by (\ref{Gf_VM_OU}) but  with $N$ instead of $N_{\rm eff}$ on the
right hand side. This crucial difference means that for the LD the diffusion parameter of
(\ref{Gf_VM_OU})  reads $2\rho^* (1-\rho^*)/N$ and is therefore
independent of the graph's structure. The OUP drift parameter is still $\tilde{s}\rho^* (1-\rho^*)$
and therefore coincides  with that of the VM  update rule. Proceeding as above, we readily find that
on average the deviations from the metastable state still vanish exponentially in time, as
$\langle \xi(t)\rangle =\xi(0)e^{-\tilde{s}\rho^* (1-\rho^*)t}$.
We can also compute the second moment of $\xi$: $\langle \xi^2(t)\rangle=
(1-e^{-2\tilde{s}\rho^* (1-\rho^*)t})/(\tilde{s}N)\sim 1/N$. Thus, the variance of the number of cooperators in the metastable
state, ${\rm var}(N\rho)=N^2(\langle \rho^2(t)\rangle - \langle \rho(t)\rangle^2)\approx N^2\langle \xi^2(t)\rangle \sim N$, scales similarly as on a complete graph.
In stark contrast with the VM case on scale-free graphs with highly-connected nodes
(when $2<\nu<3$), we thus find that the typical fluctuations in the LD case are not affected by  the  graph's structure.

\section{Fixation properties}

  \begin{figure}
\includegraphics[width=0.9\linewidth]{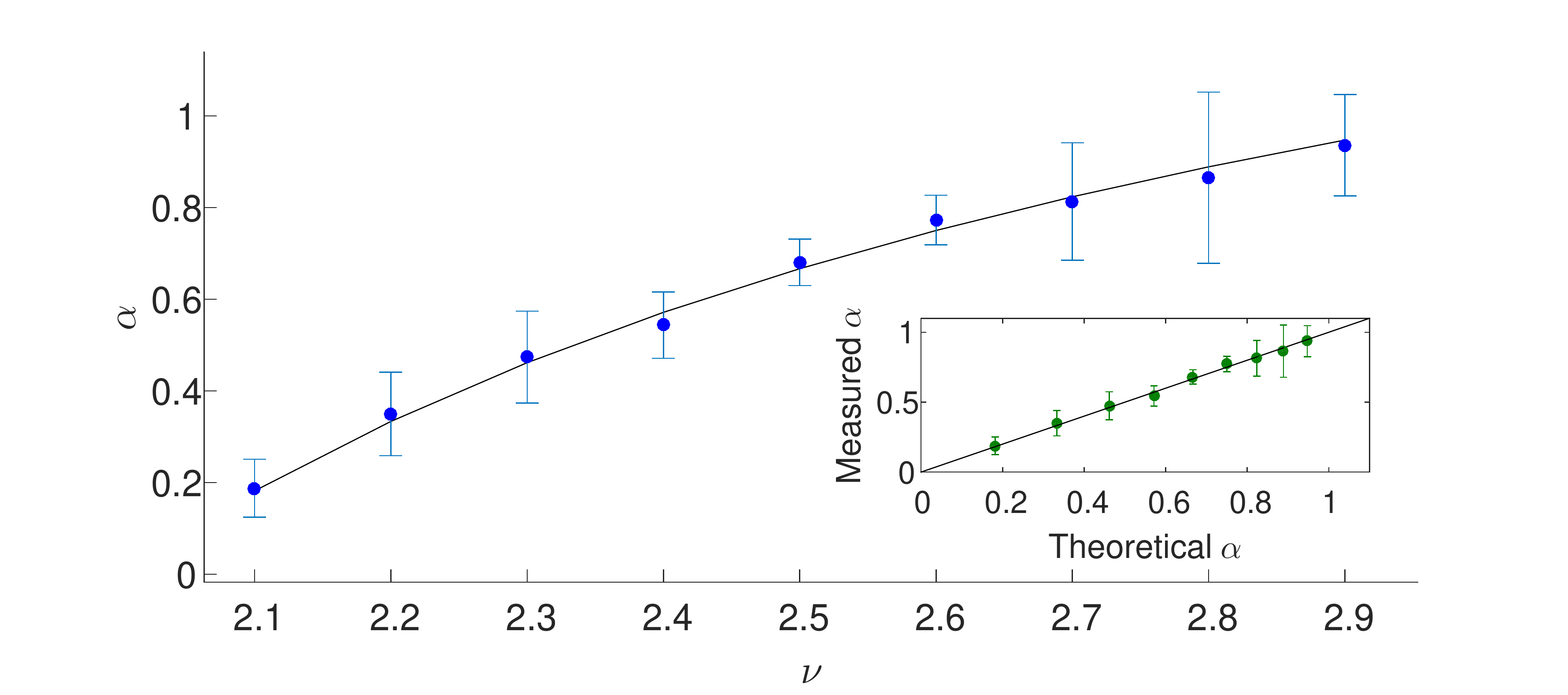}
\caption
 {(Color online).  Stretched exponential dependence of the MFT, $\ln\left(T_{\rm fix}\right)\sim N^{\alpha}$,
when the dynamics is implemented according to the VM. Here we plot the exponent $\alpha$
as a function of the exponent $\nu$ of the degree distribution.
The solid line is theoretical prediction (\ref{alpha})
giving $\alpha=2(\nu-2)/(\nu-1)$ while the symbols (with error bars)
are obtained from the simulations.
The ranges of the $N$ and $s$ parameters which were used in the
simulations are $N\in [2\cdot 10^{4},1.6\cdot 10^{5}]$
and $s\in [5\cdot 10^{-3},10^{-2}]$,  see  Appendix~\ref{app:AppendixB}. The initial conditions and payoff matrix parameters (\ref{payoffM}) are the same as in Fig.~\ref{FigMean}. Inset:
$\alpha$ obtained from simulations versus the theoretical prediction (\ref{alpha}), where the line $y=x$ is a guide for the eye.}
 \label{FigMFT}
\end{figure}

The Moran-like evolutionary processes that we are considering are absorbing Markov chains and their fate is to reach one of the
states corresponding to the entire network being populated only by cooperators or by defectors. Two important quantities to characterise the underlying evolutionary dynamics are therefore the
(unconditional) mean fixation time (MFT), $T_{\rm fix}(\rho)$, which is the mean time
to reach either of the absorbing states from an initial density $\rho$ of cooperators,
and the fixation probability $\phi^C(\rho)$ that cooperation prevails (the final state is all-$\textsf{C}$, with the extinction of all $\textsf{D}$'s)
starting from a fraction $\rho$ of cooperators.
 It is well known that on complete graphs, for ACGs these quantities scale exponentially with the
population size: $\ln{T_{\rm fix}(\rho)}\sim sN$ and
$\ln{\phi^C(\rho)}\sim -sN$ or $\ln{(1-\phi^C(\rho))}\sim -sN$, see {\it e.g.}~\cite{Antal06,MA10,AM10}.
Here, we use the effective diffusion theory to compute  $T_{\rm fix}(\rho)$ and
$\phi^C(\rho) $ on scale-free graphs and, together with large-scale computer simulations,
to uncover under which circumstances, the complex topology affects the MFT.

While we are mainly interested in the MFT, it is useful to start our discussion by outlining the calculation of the fixation
probability.
In the realm of the effective diffusion approximation, by assuming that fixation occurs after
lingering for a long time in the metastable state (see Figs.~\ref{Fig1} and
\ref{Fig2}), we use the backward operators (\ref{Gb_eff}) and (\ref{Gf_eff_LD})
to compute  the fixation probability which satisfies
$\widetilde{{\cal G}}_{ \rm b}(\rho)\phi^C(\rho)=0$  with
 boundary conditions
$\phi^C(0)=1-\phi^C(1)=0$~\cite{weaksel1,weaksel2,weaksel3,Gardiner}, and by now approximating $\omega\approx \rho$ in (\ref{Gb_eff}). This yields
%
$\phi^C(\rho)=\frac{
{\rm erfi}\left[\rho^*\sqrt{\sigma}\,\right] -
{\rm erfi}\left[(\rho^*-\rho)\sqrt{\sigma}\,\right]
}{{\rm erfi}\left[\rho^*\sqrt{\sigma}\,\right]+{\rm erfi}\left[(1 -\rho^*)\sqrt{\sigma}\,\right]}$~\cite{PRL12},
%
where ${\rm erfi}(z)\equiv\frac{2}{\sqrt{\pi}}\int_{0}^z e^{u^2} du$ and
\begin{eqnarray}
 \label{sigma}
\sigma=\left\{ \begin{array}{ll}
         \tilde{s}N\frac{\mu_1^2}{\mu_2}, &  \text{(for the VM)},\\
       \tilde{s}N, &  \text{(for the LD)}.
               \end{array} \right.
\end{eqnarray}
This clearly indicates that the VM/LD lead to very different fixation properties on scale-free networks with nodes of high degree: For the VM,
the topology yields an effective population size $N_{\rm eff}$
(\ref{Neff}) leading to $\sigma\ll  \tilde{s}N$ when $2<\nu<3$. The dependence of $\phi^C$ on the system size
when $2<\nu<3$ is thus a stretched exponential with exponent $-N^{\alpha}$ and $\alpha<1$, whereas  the fixation
probability with the LD (and for the VM with $\nu>3$) is the same
as on a complete graph of size $N$.

\begin{figure}
\includegraphics[width=0.9\linewidth]{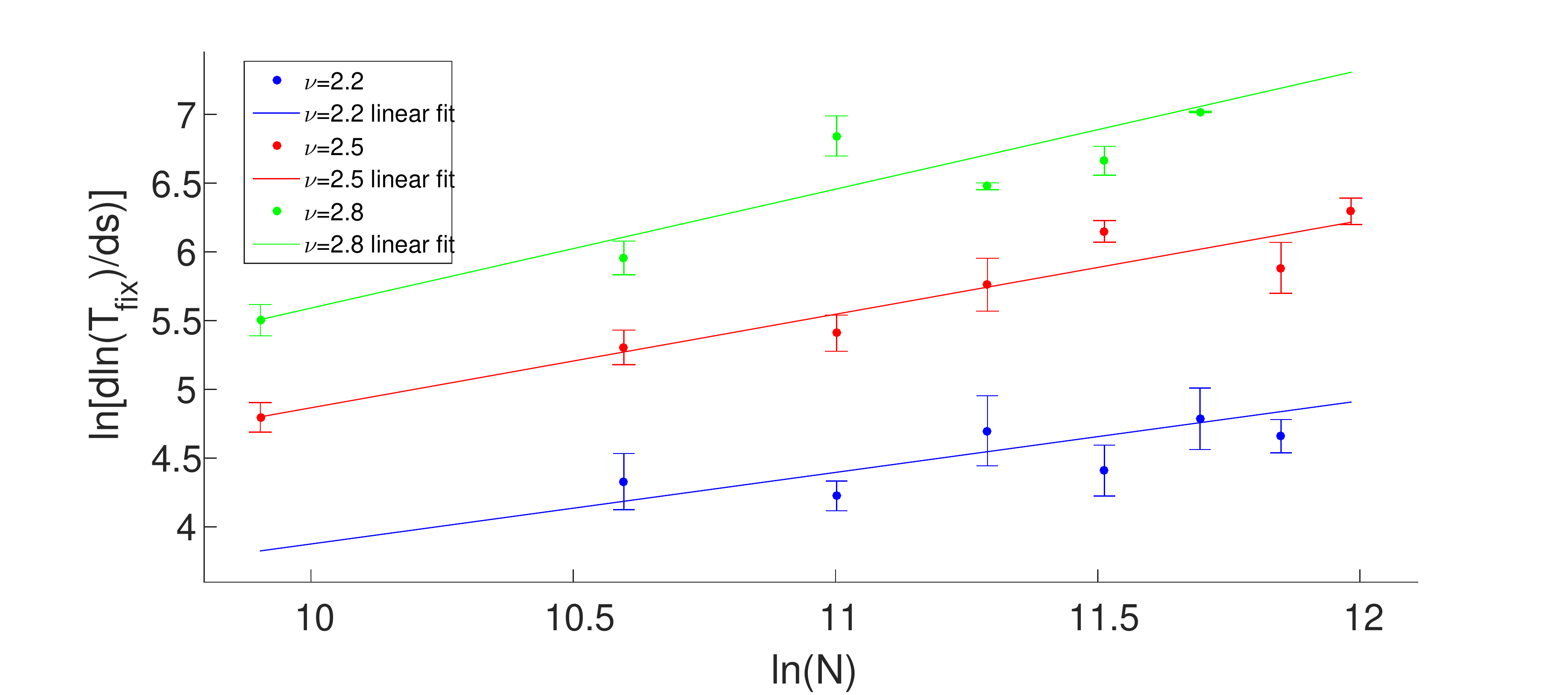}
\caption
 {(Color online). $\ln\left(d\ln\left(T_{{\rm fix}}\right)/ds\right)$
  when the dynamics is
 implemented according to the VM. Results are shown for different values of
the exponent $\nu$ of the power-law degree distribution, see legend. The range of the $s$ values
which were used here was $s\in [5\cdot10^{-3},10^{-2}]$, see  Appendix~\ref{app:AppendixB}.
The initial conditions and payoff matrix parameters (\ref{payoffM}) are the same as in Fig.~\ref{FigMean}.
Data were linearly fitted for each $\nu$
giving the slopes:
$0.3497\pm 0.0632$ (for $\nu=2.2$), $0.6805\pm 0.0509$ (for $\nu=2.5$),
$0.8651\pm 0.1867$ (for $\nu=2.8$). The corresponding theoretical slopes [Eq.~(\ref{alpha})] are
$0.3333$, $0.6667$ and $0.8889$.
}
 \label{FigMFTslope}
\end{figure}

In addition to the fixation probability, we can similarly compute the MFT, $T_{\rm fix}(\rho)$. This is done by
solving $\widetilde{{\cal G}}_{ \rm b}(\rho)T_{\rm fix}(\rho)=-1$  with
 boundary conditions $T_{\rm fix}(0)=T_{\rm fix}(1)=0$~\cite{Gardiner}. Using standard methods~\cite{Kimura,Ewens,Gardiner}, the solution to
this inhomogeneous backward FPE is given in Ref.~\cite{PRL12}
and to leading order we find:
\begin{eqnarray}
\label{MFT1}
 T_{\rm fix}(\rho)\sim
\left\{ \begin{array}{ll}
         (1-\phi^C(\rho)) e^{(\rho^*)^2\,\sigma}, &  \text{when $\rho > \rho^*$},\\
       \phi^C(\rho) e^{(1-\rho^*)^2\,\sigma}, &  \text{otherwise.}
               \end{array} \right.
\end{eqnarray}
Hence, with the expression of $\phi^C$,
when $\rho^*<1/2$  and $\rho > \rho^*$ this  gives
\begin{equation}
 \label{MFTsmall}
\ln{T_{\rm fix}(\rho)}\simeq (\rho^*)^2~\sigma.
\end{equation}

This shows that when  initially the fraction of cooperators is not too low,  metastability occurs
 prior to fixation  and the leading contribution to the MFT~(\ref{MFTsmall}) is independent of the initial condition~\cite{AM1,AM2,AM3,AM4,MA10,AM10,AM16}. It stems from this result that fixation occurs much more rapidly for the VM
 on  scale-free networks with $2<\nu<3$ than with the LD. In the VM case, the MFT grows as the stretched
 exponential $\ln{T_{\rm fix}}\sim N^{\alpha}\ll N$ with the exponent $\alpha$ given by (\ref{alpha}).
 This theoretical prediction is confirmed by our stochastic simulations (within error bars) from which we have computed the exponent $\alpha$ reported in Fig.~\ref{FigMFT} (see Appendix \ref{app:AppendixB} for technical details). The results reported in Fig.~\ref{FigMFTslope}
 corroborate, within error bars, the theoretical prediction (\ref{MFTsmall}),
  $\ln\left(d\ln\left(T_{{\rm fix}}\right)/ds\right)\sim [2(\nu-2)/(\nu-1)] \ln{N}$
 when $2<\nu<3$.  A similar analysis with LD, reported in Fig.~\ref{FigMFTslopeLD}, see Appendix \ref{app:AppendixB}, confirms
 our theoretical result (\ref{MFTsmall}) that the MFT of ACGs evolving with the LD
 grows purely exponentially with $N$ as on complete graphs~\cite{Antal06,MA10,AM10}.

 We therefore emphasize that, contrary to what stated in Refs.~\cite{PRL12,AM13}, $\phi^C$ and the MFT
 of ACGs evolving with the LD are not affected by
the scale-free topology. Therefore, in ACGs with the LD, $T_{\rm fix}$
always grows exponentially with $\tilde{s}N$ (within error bars)
as illustrated in Fig.~\ref{FigMFTslopeLD}. Our analysis of the LD in Refs.~\cite{PRL12,AM13} was based on relatively small
systems (up to $N=4,000$) which apparently were governed by finite-size effects leading to a nonlinear dependence on $N$
when $2<\nu<3$. Two of us tried to explain these effects, that turn out to disappear at large-enough $N$, in terms of an effective diffusion theory derived by assuming that $\omega$
was approximately conserved~\footnote{In Ref.~\cite{AM13} the fixation probability of coordination games subject to
the LD was considered similarly.}. Even though the difference between $\omega$ and $\rho$ is negligible for $s\ll 1$ (see Fig.~\ref{Fig2}), considering $\omega$ to be constant instead of $\rho$ leads to a large error in the MFT. Here, by carrying out extensive simulations on large systems (up to two orders of magnitude larger than in
\cite{PRL12,AM13}), and by correcting our theoretical analysis, we have confirmed that the fixation properties of ACGs evolving with the LD exhibit a linear dependence on $N$ which, as seen above, is consistent with the (approximate) conservation of $\rho$ by the LD under weak selection (see Fig.~\ref{Fig2}).

\begin{figure}
\includegraphics[width=0.9\linewidth]{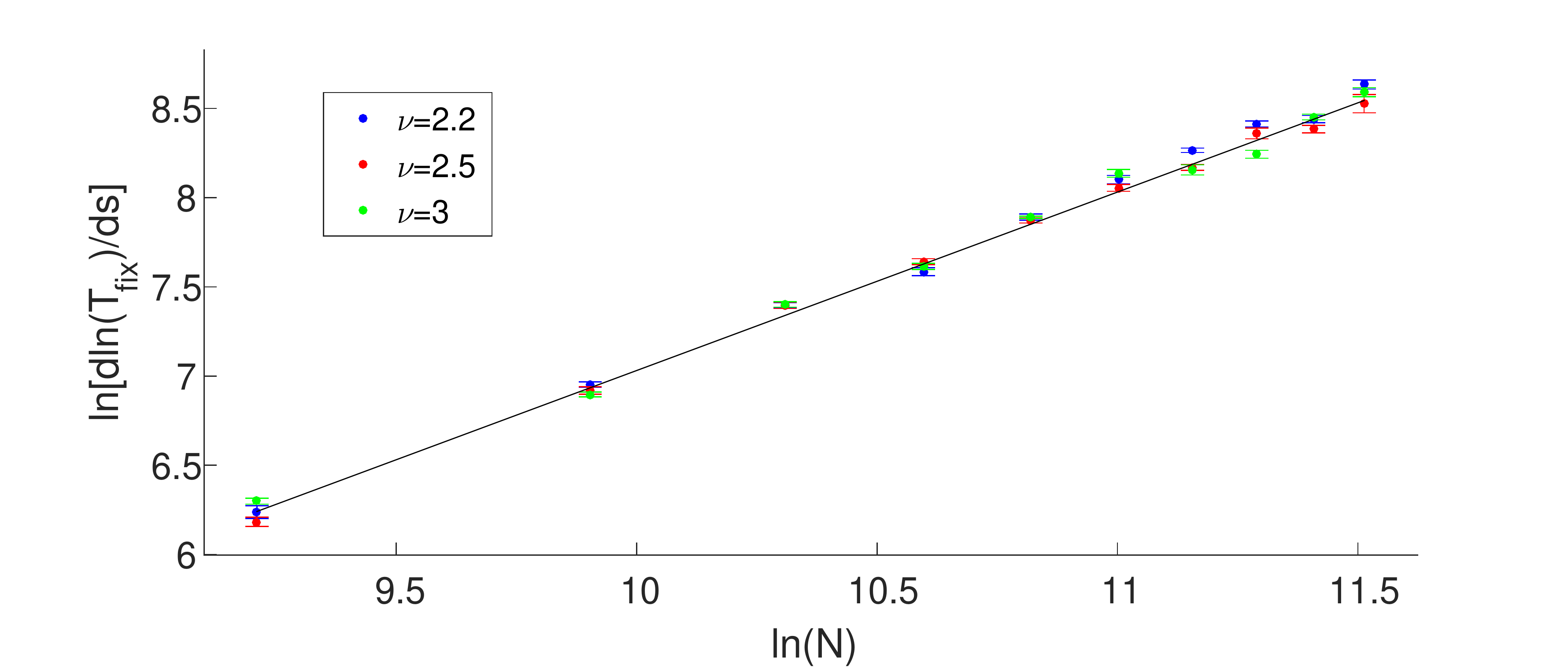}
\caption
 {(Color online). $\ln\left(d\ln\left(T_{{\rm fix}}\right)/ds\right)$
  as a function of $\ln\left(N\right)$ when the system evolves
with the LD. Results are shown for different values of
the exponent $\nu$ of the power-law degree distribution, see legend. The range of the $s$ values
which were used here was $s\in [1\cdot10^{-4},1.9\cdot10^{-3}]$, see  Appendix~\ref{app:AppendixB}.
The initial conditions and payoff matrix parameters (\ref{payoffM}) are the same as in Fig.~\ref{FigMean}.
Data for each $\nu$ were linearly fitted to get the following slopes:
$1.0316\pm0.0096$ (for $\nu=2.2$),  $1.0094\pm0.0093$ (for $\nu=2.5$),
$0.9876\pm 0.0136$  (for $\nu=3$). The line with a slope of $1$ is plotted as a guide for the eye.}
 \label{FigMFTslopeLD}
\end{figure}

\section{Summary \& Conclusion}

In this work we have studied the interplay between scale-free topology, demographic fluctuations and frequency-dependent
selection on the dynamics of anti-coordination games.
This  class of paradigmatic games, particularly relevant in theoretical biology,
 between two competing types (cooperators and defectors) is characterized by their long-lived coexistence, and eventually by the
fixation of one type. These features are here investigated  by combining  analytical methods and extensive stochastic simulations.
Since evolutionary dynamics on heterogeneous graphs is known to depend on the  underlying microscopic update rule,
we have investigated two types of  individual-based dynamics: according to the
death-first/birth-second voter model
(VM), or according to the  links dynamics (LD) in which birth/death or death/birth events occur
randomly between connected agents.  While we here specifically focus on scale-free graphs, it is noteworthy that our approach is valid for
any degree-heterogeneous networks.

Our analytical approach, valid in the weak selection limit, is based on a diffusion approximation  for the density of cooperators at
nodes of a prescribed degree. The resulting multi-variate Fokker-Planck equations are greatly  simplified by
exploiting a timescale separation: After a transient, when the selection pressure is weak, the multi-body dynamics towards fixation
can be described in  terms of the (approximately conserved) cooperator degree-weighted density (for the VM) and the  cooperator density (for the LD). As a result, the typical fluctuations in the number of cooperators in the coexistence metastable state, and
the fixation properties can be determined by means of effective single-variate forward and backward Fokker-Planck equations.

In particular, with the VM update rule the complex scale-free topology is responsible for an effective
reduction of the population size $N$ when $2<\nu<3$, into $N_{{\rm eff}}=N^{\alpha}$ with
an exponent $0<\alpha<1$ that depends on the first two moments of
 the graph's degree distribution $n_k\sim k^{-\nu}$. This results in larger typical  fluctuations and
 a superlinear dependence on $N$ of the variance of the
 number of cooperators in the metastable state. As an outcome, the MFT is exponentially decreased
 and displays a non-trivial stretched-exponential dependence on the population size.
 In the absence of ``hubs'' ($\nu>3$), in the leading exponential order, the dynamics is
 independent on the scale-free topology and the MFT scales linearly with $N$ as in the well-mixed dynamics.

When the system evolves according to the LD, the scale-free topology does not effectively renormalize the population size
 and we find $ N_{{\rm eff}}=N$ (for $\nu>2$). With the LD, the variance of the number of cooperators in the
 metastable state grows linearly with $N$ and the MFT displays a pure exponential dependence on $N$ as shown
 in Fig.~\ref{FigMFTslopeLD}, similarly to the well-mixed case. These results are in contrast with what was
 reported in \cite{PRL12,AM13}. The correction here was made possible due to a subtle amendment that we have made in
 the theoretical description accompanied by an efficient numerical code that allowed us to attain
 very large system sizes and hence to carefully analyze the dependence of the fixation properties on $N$.

\section{Acknowledgments}
DS and MA were supported by Grant No. 300/14 of the Israel Science
Foundation. MM is grateful for the hospitality of the LSFrey
at the Arnold Sommerfeld Center (University of Munich) where part of this work was done,
as well as for the financial support of the Alexander von
Humboldt Foundation by Grant No. GBR/1119205 STP.

\appendix

\section{Scale-free networks generation}

\label{app:AppendixA}

\begin{figure}
\includegraphics[width=0.9\linewidth]{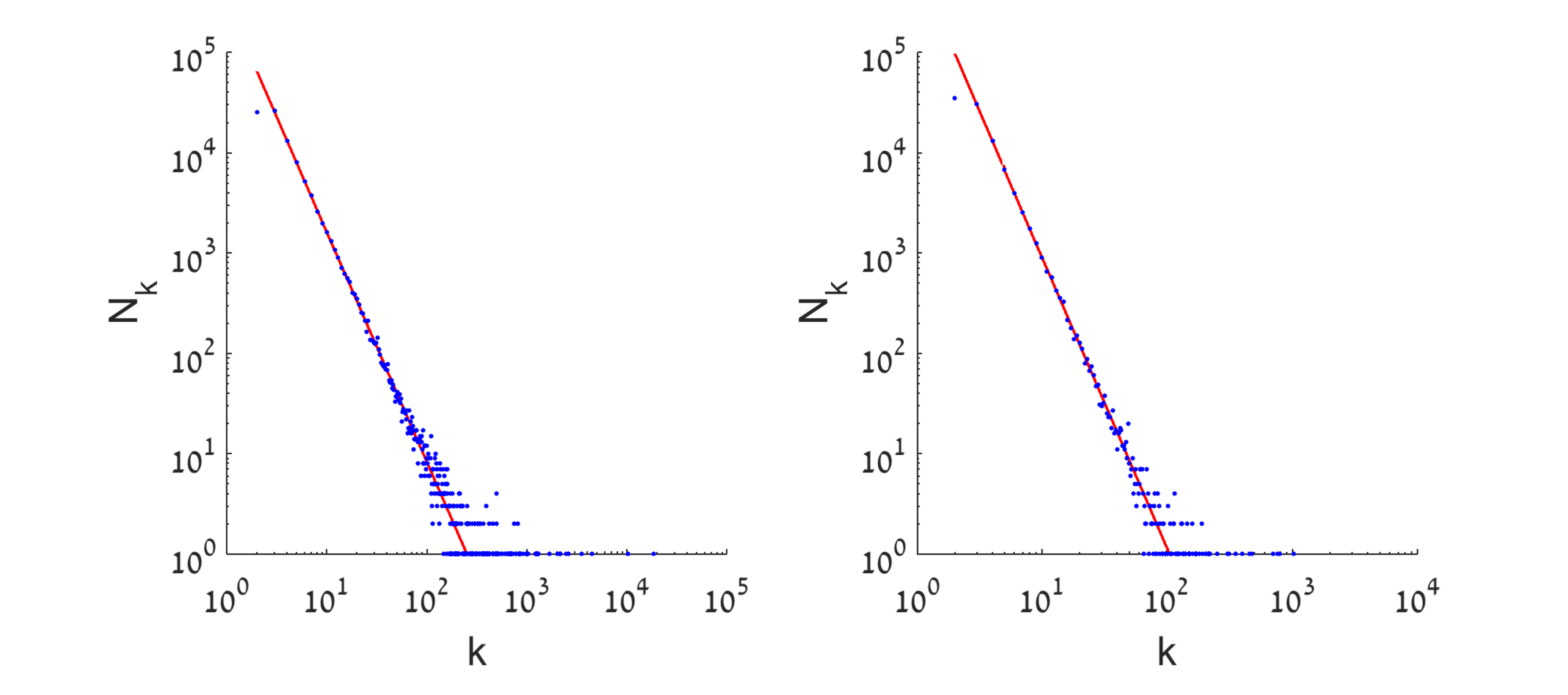}
\caption{Histogram of the number
$N_{k}=Nn_k$ of nodes of degree $k$ in a scale-free graph generated by the method of Appendix~\ref{app:AppendixA}.
Here we plot $N_{k}$ for two typical scale-free graphs of $N=100,000$ nodes
and different values of $\nu$. In each panel there is a fit to a power-law distribution of the form $N_{k}\sim k^{-\nu}$. Left panel: expected (theoretical) slope: $\nu=2.3$, measured (fitted) slope: $\nu=2.2913$. Right
panel: expected (theoretical) slope: $\nu=2.9$, measured (fitted) slope: $\nu=2.9006$.
Note that the number of nodes $N_k$ with degree
greater than $k_{\rm max}\approx  N^{1/(\nu -1)}$ is negligible compared to $N$ (here
$k_{\rm max}\approx 7000$ for $\nu=2.3$ and
$k_{\rm max}\approx 430$ for $\nu=2.9$).}
\label{NkDist}
\end{figure}

In a scale-free network the degree distribution behaves as a power law $n_k \sim k^{-\nu}$.
Here, we describe the algorithm that we have
used to generate a scale-free network of $N$ nodes with exponent $\nu$ based on the configuration model, see \textit{e.g.},
Ref.~\cite{Configuration}.  The method to generate scale-free  graphs, as those whose unnormalized degree distribution $N_k=Nn_k$
 is illustrated in  Fig.~\ref{NkDist}, consists of the following three steps:
\begin{enumerate}
\item Each node $i$ in the network receives a random number of neighbors $k_i$ to be connected to, with $k_i\in[2,N-1]$, sampled from a power law distribution $n_k$.
Choosing a random number $R\in(0,1]$ from a uniform distribution, the power law distribution $k^{-\nu}$ is generated by using the expression $k={\rm round}\left(\left[\left(k_{+}^{1-\nu_{sim}}-
k_{-}^{1-\nu_{sim}}\right)R+k_{-}^{1-\nu_{sim}}\right]^{\frac{1}{1-\nu_{sim}}}\right)$
where $\nu_{sim}$ is the target $\nu$, $k_+=N-1$, $k_-=2$, and ${\rm round}(z)$ gives
the closest integer to $z$.

\item We create links according to the designated number of neighbors of node $i$, $k_i$.
However, a link between $i$ and $j$, $i\neq j$, is created only if both nodes' degree is lower than their corresponding $k_i$ and $k_j$. No multi-links (between $i$ and $j$) and no self-links are allowed.

\item A histogram of the number of nodes with $k$ neighbors is created. We measure the slope of the histogram in a log-log axis representation to check if it coincides with $-\nu$ within $1\%$. As this  distribution is noisy for very small/high values of $k$, we measure the slope
by taking $k$ values such that $k\ge5$ and $N_{k}>100$. When the measured slope was not within $1\%$ of the desired slope, we repeated the
process starting from Step $1$ with an increased/decreased value of $\nu_{sim}$ by $0.01$, when the slope was too low/high.
\end{enumerate}

The procedure ends when the measured slope coincided with $-\nu$ within $1\%$ tolerance, see example in Fig.~\ref{NkDist}.
It has to be noted that in this implementation, nodes are allowed to have a maximum degree
$N-1$, but in practice the number of nodes of degree greater than $k_{\rm max}\sim N^{1/(\nu -1)}$
is negligible, as can been seen in Fig.~\ref{NkDist}. Therefore, as in Sec.~V, we can safely consider
that the power-law distribution $n_k\sim k^{-\nu}$ holds up to the maximum degree, $k_{\rm max}\sim N^{1/(\nu -1)}$
(size of the largest hub)~\cite{VotNets11,VotNets12,VotNets13,KS02}. Furthermore, in this procedure the correlation between nodes of degree $k$ and their neighbors can be shown to be negligible as long as $k$ is in the bulk of the power-law distribution~\cite{Configuration} (for example in the left panel of Fig.~\ref{NkDist} one can show that correlations are negligible as long as $k<100$, which basically includes all the nodes of the network except the noisy right tail.)

\section{Computations of the cooperators mean and variance, and the MFT}
\label{app:AppendixB}
In order to corroborate our analytical results for the mean and variance of the cooperators density, as well as for the MFT, we have performed extensive computer simulations for various values of $\nu$, $N$ and $s$. Throughout the paper, we have tried to achieve a compromise between computational efficiency and numerical robustness, and therefore, we have focused on graphs with number of nodes spanning from $2\cdot 10^4$ to $2\cdot 10^5$. Indeed, as shown in Refs.~\cite{PRL12,AM13}, graphs of smaller size are strongly influenced by finite-size effects, while simulations on graphs of larger size were extremely time-costly.

We first describe how
$\left\langle N\rho\right\rangle$ and ${\rm var}\left(N\rho\right)$ have been computed in the long-lived metastable state.
In Figs.~\ref{FigMean} and ~\ref{FigVar} each point is computed by averaging over up to ten graph realizations stemming from ten
different scale-free graphs $G_i\left(N,\nu\right)$, see Appendix~\ref{app:AppendixA}. For each simulation we calculate the
mean of $\rho$ and its variance between the times $t=500$ (well above the time scale required to reach metastability, see text)
and $t=25000$, where we omit simulations which have fixated within this time window.  The error is measured by computing the
standard deviation
of the results of each realization divided by the square root of the number of realizations taken into account.

We now outline how the unconditional mean fixation time (MFT) is computed from our  simulations. In Fig.~\ref{FigMFT} we show the  stretched exponential dependence of
the MFT, $\ln\left(T_{\rm fix}\right)\sim N^{\alpha}$,
when the dynamics is implemented with the VM and $2<\nu<3$. The computation
of each data point of Fig.~\ref{FigMFT}  is obtained from the following procedure:\\
(i) For each set of parameters $\left(N,\nu,s\right)$ and scale-free graph realization
$G_i\left(N,\nu\right)$, we average the fixation time over up to $200$ replicas (each
with a different initial condition). That is, given eight different graph realizations, we run a total of
1,600 replicas for each $N,\nu, s$. The averaging is done by fitting the fixation times histogram of the replicas of the same graph by an
exponential distribution, and by multiplying  by $N$ (time steps are divided by $N$). \\
(ii) Having found the MFT, $T_{\rm fix}$, for given $N$, $s$, $\nu$ and graph $G_i$, we fit the logarithm of the MFT as a function of $s$, for given $N$, $\nu$ and graph $G_i$. Since we theoretically expect that $\ln T_{\rm fix}\sim sN^{\alpha}$, the slope of this fit, $d\ln T_{\rm fix}/ds$ gives us the dependence of $\ln T_{\rm fix}$ on $N^{\alpha}$. \\
(iii) Having found the slopes as function of $N$ for given $\nu$ and graph $G_i$, we average over the slopes  stemming
from the different graphs, which yields a single slope depending on $N$ and $\nu$.
Here we have only taken the results from those graphs, where the slope as a function of $s$, computed from the lower
values of $s$, was consistent (up to factor $2$) with that of the higher values of $s$. \\
(iv) We now fit the logarithm of this averaged slope, $\ln (d\ln T_{\rm fix}/ds)$ as a function of $N$ to find the exponent
$\alpha$ as function of $\nu$. In Figs.~\ref{FigMFTslope} and \ref{FigMFTslopeLD} we plot $\ln (d\ln T_{\rm fix}/ds)$ as
function of $N$, for various values of $\nu$. While in the VM  we find that the averaged slope strongly depends on $\nu$, in
the case of LD the slope is independent on $\nu$, see text. In these figures the error bars originate from the fitting step (ii) and from
averaging over the different graphs.\\
(v) Finally, $\alpha$  is plotted as a function of $\nu$ in Fig.~\ref{FigMFT}. Here, the error bars for each $\nu$ are computed from the linear fit described in step (iv).

\end{document}